# THRESHOLD FOR EXTENDED EMISSION
# IN SHORT GAMMA-RAY BURSTS

Jay P. Norris[1], Neil Gehrels[2], and Jeffrey D. Scargle[3]

[1] Physics and Astronomy Department
University of Denver, Denver CO 80208

[2] Astroparticle Physics Laboratory
NASA/Goddard Space Flight Center, Greenbelt, MD 20771

[3] Space Science and Astrobiology Division,
NASA/Ames Research Center, Moffett Field, CA 94035-1000.

## ABSTRACT

The initial pulse complex (IPC) in short gamma-ray bursts is sometimes accompanied by a softer, low-intensity extended emission (EE) component. In cases where such a component is not observed, it is not clear if it is present but below the detection threshold. Using Bayesian Block (BB) methods, we measure the EE component and show that it is present in one quarter of a *Swift*/BAT sample of 51 short bursts, as was found for the *Compton*/BATSE sample. We simulate bursts with EE to calibrate the BAT threshold for EE detection and show that this component would have been detected in nearly half of BAT short bursts if it were present, to intensities $\sim 10^{-2}$ counts cm$^{-2}$ s$^{-1}$, a factor of five lower than actually observed in short bursts. In the BAT sample the ratio of average EE intensity to IPC peak intensity, $R_{int}$, ranges over a factor of 25, $R_{int} \sim 3\times10^{-3} - 8\times10^{-2}$. In comparison, for the average of the 39 bursts without an EE component, the 2-$\sigma$ upper limit is $R_{int} < 8 \times 10^{-4}$. These results suggest that a physical threshold effect operates near $R_{int} \sim$ few $\times\ 10^{-3}$, below which the EE component is not manifest.

Subject headings: gamma-ray burst: general

# 1. INTRODUCTION

As described in Norris & Bonnell (2006), inspection of *Compton*/BATSE bursts during the *Compton* mission suggested an apparent (sub)class of bursts with a short initial pulse complex (IPC) followed by a low-level extended emission (EE) component, but the character and significance of this burst subset were not clear until first detections of similar bursts by *Swift* and HETE-2 (Barthelmy et al. 2005, Villasenor et al. 2005). The main issues included how to discern unambiguously low-intensity EE in the presence of background variations for BATSE, as well as the fact that short bursts with EE complicated burst classification. Norris & Bonnell visually recognized an evident EE component in only ~2% of the BATSE short burst sample; yet spectral lags for the IPC in BATSE and *Swift*/BAT bursts with and without EE were shown to be comparable, consistent with negligible lag. Then in an objective Bayesian Block treatment, a significant EE component was shown to be present in one quarter (~ 64 / 256 bursts) of a BATSE short burst sample pruned to retain bursts with low background variation (Norris & Gehrels 2009). Also, the loci of fundamental burst parameters – duration and spectral hardness (Kouveliotou et al. 1993) – were found to be indistinguishable for the two BATSE subsets of short bursts (Norris & Gehrels; Figure 2).

Rhoads (2008) framed the problem succinctly in the modeling context, exploring whether two progenitor types, or a continuum of inspiral types, might explain short bursts with and without the EE component. The environmental evidences presented so far are, at least, contrary: The analysis in Troja et al. (2008) showed that short burst sources with only an IPC lie 10's of kpc from their hosts, whereas the IPC+EE sources lie significantly closer to their hosts, a few kpc, with the different distance scales qualitatively commensurate with NS-NS and NS-BH merger timescales, respectively. Yet Fong, Berger & Fox (2009) in an HST survey of short GRB hosts found no evidence for the dichotomy suggested by Troja et al.; however, they did find that offsets are indeed greater than for long GRBs, but only in proportion to their larger host sizes. Also supportive of the dual progenitor view, Sakamoto & Gehrels (2009) found that IPC-only bursts have short-lived X-ray afterglows ($< 10^4$ s), no optical afterglows, and a relatively low-intensity IPC. But this finding is challenged by the results of Nysewander, Fruchter & Pe'er (2009), who find that long and short bursts' optical and X-ray afterglows are similar.

There are even motivations and discussions to classify some short bursts so that they might fit into the generic core collapse scenario, e.g., in terms of prompt+environmental characterizations (Zhang et al. 2009), and due to detection of apparent, higher-redshift short bursts like GRB 090426 with z = 2.609 (Levesque et al. 2009). Still, the most recent, extensive analysis of short burst afterglows and their hosts by Kann et al. (2008) shows that their optical afterglows have a significantly lower average luminosity than those of long bursts; and never any evidence supporting a supernova event, as would be expected for core collapse progenitors (also Kocevski et al. 2009; GRB070724A: no evidence for SN).

The current picture of short bursts with and without an EE component would be partially clarified if it could be shown that a dichotomous classification concerning the prompt emission obtains, in the sense that some short bursts do not have an EE component that remains undetected at a level below the instrument sensitivity. In this work we demonstrate that the BAT instrument is sensitive to the EE component to significantly lower fluxes than it is manifest in the BAT sample. In the next section we describe the particulars of time series representation and analysis of the BAT sample of short bursts using a Bayesian Block methodology (Scargle 1998). The subsequent section describes the results gauged in terms of BAT sensitivity to the low-intensity EE. The mask-tagged background-subtracted data type – available by virtue of BAT's coded aperture / pixelated detector combination – is particularly amenable to the BB approach, and yields sensitive limits for the EE component. In the discussion section we speculate on possible implications for physical mechanisms of IPC and EE, assuming a dichotomous classification for short bursts. In the Appendix we present the formalism for BB analysis of Gaussianly distributed data – that is, the particular form of the fitness function – and a method for adjustment of the BB prior for division of blocks.

## 2. BAYESIAN BLOCK ANALYSIS

We describe the BAT mask-tagged data types, construction of the burst time series, the Bayesian Block procedures used to measure IPC and EE properties, and calibration of the BAT sensitivity to the presence of an EE component.

### 2.1 *Time Series Representation*

The background level for the BAT full array is ~ 6,000–9,000 counts $s^{-1}$ (15–350 keV), of the same order of magnitude as that for GRBs detected by BATSE. However, a major difference with BATSE is that short and long timescale background fluctuations, arising from other sources in the BAT field of view, are virtually absent for the mask-tagged background-subtracted data types (Fenimore et al. 2003). Briefly, the mask-tagging procedure effectively sums to zero those counts in pixels from directions other than the burst direction. This is performed by weighting photons recorded in a given pixel (i.e., one of BAT's 32K 4 mm square detectors) by +/– 1, according to whether the pixel is more/less than 50% illuminated by the coded aperture as seen from the refined burst direction (e.g., the XRT position). The signal from the burst direction is diminished in the process, *but the resulting sky background is essentially zero*. The mask-tagged data type produced by this procedure comprises a rate array and an auxiliary array with approximately Gaussianly distributed errors.

For this analysis we used the BAT 1-s and 64-ms, 4-channel data types, available at a *Swift* Data Center web site[1] as coded by trigger number (e.g., trigger 412217 ≡ GRB 100213A). The native units of the rate and associated error arrays are counts pixel$^{-1}$ s$^{-1}$. The EE component, while softer than the IPC, contributes significantly in the highest BAT channel. Therefore we summed over the 4-channel data, producing for each burst a 15–350 keV time series.

To measure more accurately the IPC duration and peak intensity we replaced the 1-s data, 6 s prior and 10 s after the burst trigger, with 64-ms data, appropriately apportioning counts and errors in the two conjoining bins. Each constructed time series extended a maximum of 200 s (400 s) before (after) burst trigger. For eight bursts the interval before trigger was shorter, ranging 60–180 s. The available intervals post trigger were routinely 300 s for GRBs 060313 and earlier. For six bursts shorter post trigger intervals were available: 240, 170, 230, 180, 240, and 190 s, for GRBs 050202, 050925, 070923, 071112B, 090510, and 090815C, respectively. (These intervals are ~ twice as long as the average detected EE duration; no EE component was detected for these six bursts.)

## 2.2 *Bayesian Block Analysis of Short Burst Time Series*

The EE component is usually much lower in intensity compared to the IPC; for most short bursts with EE it is lower by a factor of $10^{-3} - 10^{-2}$. The EE intensity is often comparable to the error in the background rate, ~ $10^{-2}$ counts pixel$^{-1}$ s$^{-1}$, and therefore detectable only because its characteristic timescale is $\tau_{EE}$ ~ 100 s, roughly two orders of magnitude longer than the IPC. Thus, accurate knowledge of background level is a paramount requirement in this analysis. As discussed above, the BAT mask-tagged data robustly satisfies this requirement.

It remains to characterize the IPC and EE intensities and durations in the presence of the Gaussianly distributed (signal +) background errors. We utilize Bayesian Block (BB) methods (Scargle 1998), which are objective and automatic, do not rely on pre-defined bins or thresholds, optimally reveal signal but suppress noise, based on a sound statistical procedure, and – particularly important for this work – provide time-profile features convenient for further analysis and interpretation. The compression of a continuous signal into discontinuous blocks, despite appearances, preserves most or all of the information present in the data.

In the minimalist "constant mean" model, the BB algorithm divides a time series into blocks whose intensity over the block interval equals the average of counts within that interval. The action at the "atomic level" of the BB algorithm determines whether and where to segment an interval (a block) into two intervals (two blocks), assuming only a constant mean over each interval. The optimal algorithm for a given time series is global: the final blocking is evaluated holistically across the entire time series. The algorithm was constructed as an example of Generalized Dynamic Programming (Jackson et al. 2003).

---

[1] http://swift.gsfc.nasa.gov/SDC/data/local/data1/data/old_products/00412217000/bat/

Where to make divisions into blocks requires evaluation of a fitness function. The fitness function for BAT's mask-tagged, Gaussianly distributed data is described in the Appendix. (This fitness function was originally derived for this BAT data type, for the purpose of facilitating the measurement of GCN-reported BAT burst durations.)

The BB algorithm has one (apparently) free parameter, $ncp_{prior}$, the adopted prior on the number of blocks which affects the relative probability of any partition into blocks, e.g., the decision to divide one block into two. A "Change Point" is the point joining two blocks, where the constant mean model changes value. The probability against each block division is modified by the value of $ncp_{prior}$. We adjusted $ncp_{prior}$, requiring the probability for the BB algorithm to generate a "false positive" – an extraneous block in a 600-s time series like those of short bursts analyzed here – to be $\lesssim 0.02$ (~ $2\sigma$ confidence), thereby eliminating this final free parameter. The procedure for estimating $ncp_{prior}$ for the BAT short burst time profiles is also described in the Appendix.

The peak of the BB representation in the 64-ms domain of the time series was taken as the IPC peak intensity, $IPC_{pk}$. We utilize a duration measure, $T_{\{1/e\}}$ – the interval between the outermost points of $(1/e) \times IPC_{pk}$. $T_{\{1/e\}}$ is then the interval between start time of the first block and end time of the last block satisfying this condition. We have performed simulations that show this measure of duration to be relatively resilient against brightness bias. (Some granularity is introduced into the measurements of $T_{\{1/e\}}$ and $IPC_{pk}$ by the 64-ms resolution data for bursts with $T_{\{1/e\}} \lesssim 0.2$; however, this is of negligible import to our central thrust of characterizing the BAT sensitivity to the EE component.) $T_{\{1/e\}}$ has an additional merit for our particular purpose: Often a clear intensity minimum occurs between the IPC and EE components, but often not. In the former cases, we define the start of the EE component as the mid-point of the block representing the minimum; in the latter cases the end of the last block of the $T_{\{1/e\}}$ interval is selected as the start of EE. Since $T_{90}$ is a fluence measure, this approach obviously would not be possible with a $T_{90}$ characterization of the burst duration. Examples of IPC and EE (with or without an interjacent intensity minimum) in short bursts are illustrated in the Appendix, and in Figure 2 and Figure 7 of Norris & Bonnell (2006).

If after a wait period $\Delta T_{IPC \to EE}$ following the trigger time, a transition from a higher to lower intensity (background level) final block does not occur by the end of the time series, no EE component was declared to be detected. In these non-EE cases, a single block then describes the interval after the prescribed $\Delta T_{IPC \to EE}$ wait period. However, if such a downward transition does occur, then the start and end times of the transition block(s) denote the EE component interval. The $\Delta T_{IPC \to EE}$ interval was chosen as 7 s, which often corresponded to a point near the middle of the minimum between IPC and EE components, if in fact a local minimum block occurred.

For one burst, GRB 090510, a 6-s block *started and ended* during the $\Delta T_{IPC \to EE}$ wait period, with an intensity ~ 0.3 counts cm$^{-2}$ s$^{-1}$, almost two orders of magnitude less intense than the burst's IPC intensity, but more intense than most detected EE components in this study. GRB 090510 was the brightest short burst so far detected by the *Fermi*/LAT experiment (Abdo et al. 2009). We note it as an unusual case – the 6-s block may represent a very short EE component, or perhaps the tail of the IPC. The latter possibility may be favored since this 6-s interval has a spectral hardness comparable to that of the IPC, and significantly higher than the average hardness of EE components detected by the BAT.

The BAT sample of 51 short bursts analyzed here is listed in Table 1, along with the measured IPC and EE properties. After the burst date the columns are: IPC duration ($T_{\{1/e\}}$), peak intensity and associated error ($IPC_{pk}$ and $\varepsilon_{IPC}$), EE duration ($T_{EE}$), average EE intensity and associated error ($EE_{avg}$ and $\varepsilon_{EE}$), and absolute value of the intensity ratio, $|EE_{avg} / IPC_{pk}| \equiv R_{int}$. Intensity values are reported in counts cm$^{-2}$ s$^{-1}$. The conversion factor from {counts pixel$^{-1}$ s$^{-1}$} to {counts cm$^{-2}$ s$^{-1}$} is 6.25. If the EE component is detected – as in 12 of 51 bursts – a value for $T_{EE}$ is recorded in the Table. When no EE component is detected, the (~ 300–400 s) block stretching from the cessation of the IPC component to the end of the time series will have a relatively small positive or negative intensity. For the 39 cases where there is no record for $T_{EE}$, the intensity of this last block is still recorded in the $EE_{avg}$ column in Table 1 and a value (an upper limit) for $R_{int}$ is calculated from it. The last column in Table 1 indicates whether or not the *Swift* XRT detected a fading X-ray afterglow, often the best evidence that a (faint) short burst is real. Lack of detection may be attributable to: Sun or Moon constraint, or a spacecraft mode, that disallows a prompt spacecraft slew (*Swift* slews after the BAT trigger so that the pointed instruments may acquire the burst position); or, absence of a detectable afterglow even upon acquisition of the burst position by the XRT. Notable in Table 1 are GRBs 050911 and 090715A, for which EE was detected, but no X-ray afterglow reported; in both cases a prompt slew was not possible. Not included in our sample are two short GRBs for which there was no mask-tagged data generated (GRBs 070406 and 080121), and GRBs 080123 and 090417A, for which the IPC was too weak to be detected by the BB algorithm (without lowering ncp$_{prior}$ excessively, thereby incurring many extraneous false positive blocks in the BB representation).

### 2.3 *Simulations of EE component*

To calibrate the sensitivity of the BB representation to an EE signal, we performed two types of simulations, with temporal structure similar to that of the twelve bursts with EE in Table 1.

For the first set of simulations, the two aspects of the actual 51 burst time series that were preserved were the IPC component and the duration of each time series. That portion of the time series following the IPC was replaced with two intervals with 1-s binning. The first interval was fixed at 100 s, the median observed timescale for the EE component, and the second interval spanned the remainder of the time series, typically 200–300 s. The first (second) interval was populated with a single positive constant (zero) mean with Gaussianly distributed errors added. The 1-$\sigma$ errors were taken bin for bin from the actual mask-tagged data files. Many simulation sets were performed with discrete values of the constant mean in the first interval ranging from ~ 0.07 to 0.005 counts cm$^{-2}$ s$^{-1}$. The objective of this first type of simulation was to estimate the intensity levels at which a flat, 100-s EE component would be detected ~ 99%, 50% and 1% of the time. The corresponding estimated levels are 0.050, 0.025, and 0.008 counts cm$^{-2}$ s$^{-1}$.

The second type of simulations reproduced the block structure of the twelve bursts with detected EE components in Table 1, but with the block intensities distributed uniformly randomly in the logarithm between 0.050 and 0.008 counts cm$^{-2}$ s$^{-1}$ – the 99% and 1% detection points for a flat EE component. As for the first type of simulation, the blocks were rendered to 1-s binning and Gaussianly distributed errors were added. Here the objective was to estimate the fraction of "life-like" bursts across the 1–99% zone of detection probability that would have been detected *if an EE component were present*.

## 3. BAT SENSITIVITY TO EXTENDED EMISSION

In Figure 1 the IPC peak intensity (IPC$_{pk}$) is plotted versus the EE average intensity (EE$_{avg}$) for the 51 BAT short bursts in Table 1. Filled diamonds indicate the twelve bursts where an EE component was detected according to the criteria described in §2.2. For the other 39 bursts with no EE detection, open triangles (circles) indicate positive (negative) intensity values of the last block in the BB representation, typically 300–400 s, largely consistent with background level. All error bars are 1-$\sigma$, propagated from the rate errors in the native mask-tagged data (see §2). The dashed 45° lines are loci of constant R$_{int}$ ≡ |EE$_{avg}$ / IPC$_{pk}$|. The vertical solid lines at 0.050, 0.025, and 0.008 counts cm$^{-2}$ s$^{-1}$, correspond to 99%, 50%, and 1% detection efficiency, respectively, for a flat 100-s EE component. The outlier filled diamond near {IPC$_{pk}$, EE$_{avg}$} = {15, 1.2} is GRB 060614, whose borderline T$_{90}$ (also T$_{1/e}$, see Table 1) duration made for an intriguing classification debate (see Gehrels et al. 2006).

The feature of Figure 1 that we wish to further elucidate is the relatively unpopulated region between EE detections and nondetections, closely corresponding to the 99% – 1% detection efficiency zone for a 100-s flat EE component, calibrated by the first type of simulation described in §2.3. Figure 2 depicts the results from 16 typical sets of simulations of the second type, where the EE components of the twelve actual bursts have been replaced with simulated EE, distributed so that the log(EE) values uniformly populate the 99% - 1% detection probability zone. The durations of the simulated EE components are unchanged from the actual durations, as are the block patterns unchanged; only the intensities are diminished. There are 16 × 12 = 192 simulations of the EE component shown in Figure 2. The same BB analysis as described for the actual burst time series in §2.2 was performed for the 192 simulated bursts. Since the IPC component is unmodified in the eight sets, the $IPC_{pk}$ values fall on the twelve horizontal levels corresponding to those for the actual bursts. Two aspects are evident and different in Figure 2 compared to Figure 1: The EE component is detected for 87 of the 192 simulations (~ 45%), twice as frequently as for the actual bursts; and the 99% – 1% detection zone ($EE_{avg}$: 0.050 – 0.008 counts $cm^{-2}$ $s^{-1}$) is much more populated. That the simulated intensities were designed to populate this region, but some detections fall at higher intensities, is by virtue of the fact that the *detected* EE component (as represented by BBs; EE presence and duration as defined in §2.2), was sometimes shorter than ~ 100 s and of intensity higher than the EE average intensity, with the remaining undetected portion of the simulated EE interval being of lower intensity. Interestingly, this suggests that, if $\tau_{EE}$ were much shorter than 100 s in a larger fraction of the actual population of short bursts with EE observed by the BAT, we would be detecting it via the BB algorithm. Simulations described in the Appendix support this conclusion.

Thus the import of Figure 2 is that an EE component would easily be detected to intensities a factor of 4–5 lower than actually observed in short bursts, if the EE component were present. We conclude that the relatively EE-unpopulated region in Figure 1 is real, that the BAT is revealing a real physical threshold for EE in short bursts, and that ~ three fourths of BAT short bursts are truly short – not accompanied by an EE component.

We then expect that the average time series of the bursts without a detected EE component should evidence close to background level in the canonical EE interval. First, in Figure 3a we replot the 1-σ intensity upper limits for the 39 bursts without EE detections, rendered linearly rather than in the logarithm, to illustrate that the distribution of block intensities in the post-IPC interval (~ 300–400 s) is centered close to zero. Recall from §2.1 that positive or negative values occur, since pixels are weighted by +1 or –1, respectively, according to the pixel illumination fraction seen from the burst direction. In fact, significant contamination occasionally occurs due to imperfect mask-tagged background subtraction of a very bright source in the BAT field of view. This is indeed the case for the highest positive and the two most negative values plotted in Figure 3a (see Table 1 for $EE_{avg}$ values: GRBs 050202, 050925, and 060313). Course estimates of the bright-source contamination level for these three bursts can be had from pre-slew and post-slew calculations; more accurate estimates during slewing are impractical. The contamination levels in these three cases are of order ±0.01 counts $cm^{-2}$ $s^{-1}$ (T. Sakamoto, private communication), comparable to their central values in Figure 3. These are three of the four bursts with EE non-detections that inhabit the 50% – 1% detection zone in Figure 1. No such bright-source contamination occurs for the twelve bursts with EE detections. Thus, taking these three bursts into account further widens the gulf between short bursts with and without EE detections. The solid vertical line in Figure 3a represents the average of the EE upper limits for the 39 bursts, $8.2 \pm 5.5 \times 10^{-4}$ counts cm$^{-2}$ s$^{-1}$, a factor of 50 below the least intense EE detection (for GRB 051227). The dashed lines show $\pm$ 1-$\sigma$ error bars for this average upper limit – consistent with zero signal. For comparison, we also examined the pre-IPC interval (~ 100–200 s), since one might expect no signal before the trigger time in short bursts. Figure 3b shows the upper limits for the 39 non-EE bursts in the pre burst interval, again consistent with zero signal at the 2-$\sigma$ level. The pre-IPC and post-IPC block intensities are not significantly correlated, as might be expected since Swift usually slews within seconds of a burst trigger.

The average time series itself – with the 39 individual time series aligned in registration at their trigger times, and truncated to the interval {–100, +300} s – is illustrated in Figure 4 with the BB representation overplotted. The block intensity representing the average post-IPC interval (2–300 s) is $1.2 \pm 0.6 \times 10^{-3}$ counts cm$^{-2}$ s$^{-1}$. For just the 100-s interval following the IPC, the intensity of the average time series is $1.8 \pm 1.0 \times 10^{-3}$ counts cm$^{-2}$ s$^{-1}$. (We performed experiments with the average non-EE time series, decreasing ncp$_{prior}$ significantly, attempting to induce a change point in the vicinity of 100 s after the IPC – thereby incurring many false positives – but with negative outcome.) Simulations show that a 100-s EE signal in 39 co-added bursts would have been detected ~ 50% of the time at a level of ~ $5 \times 10^{-3}$ counts cm$^{-2}$ s$^{-1}$, a factor of 10 below the lowest intensity of the twelve bursts with detected EE components.

In summary, from our calibrations of the BAT sensitivity to an EE signal, and lack of a signal in the co-added bursts without an EE component, we conclude that ~ three fourths of short bursts detected by the BAT have no associated EE component – these bursts are truly short.

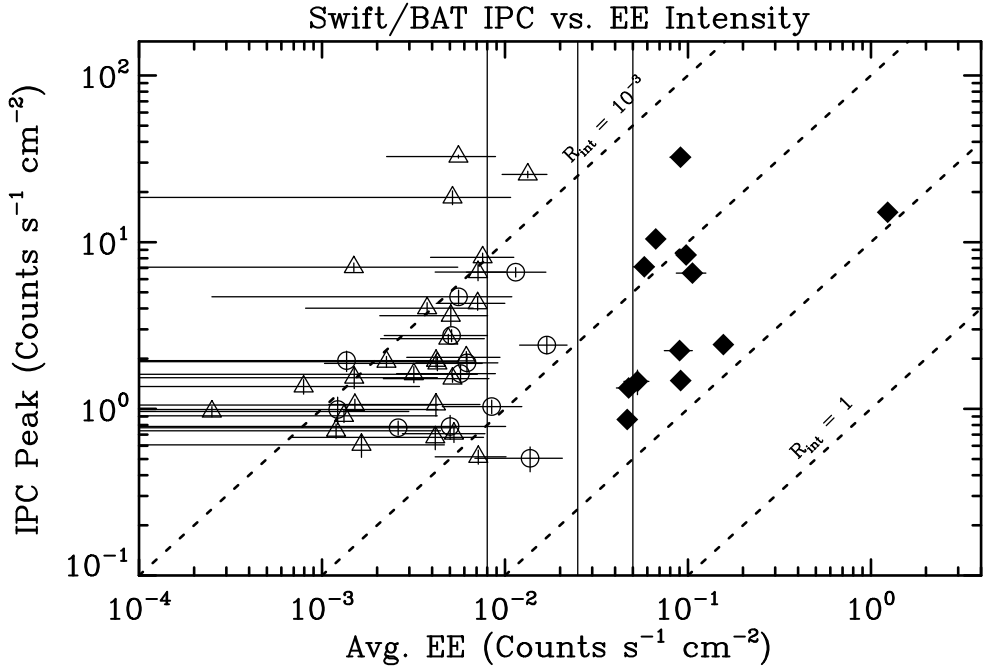

Fig. 1– IPC peak intensity (IPC$_{pk}$) vs. average EE intensity (EE$_{avg}$) for the 51 BAT short bursts in Table 1. Filled diamonds are the twelve bursts with significantly detected EE. Open triangles (circles) indicate positive (negative) upper limits for EE component, as measured by last block in the BB representation. Dashed 45° lines are loci of constant R$_{int}$ ≡ |EE$_{avg}$ / IPC$_{pk}$|. Vertical lines show (r. to l.) 99%, 50%, and 1% detection efficiency for simulated, flat 100-s EE component. All error bars are 1-$\sigma$.

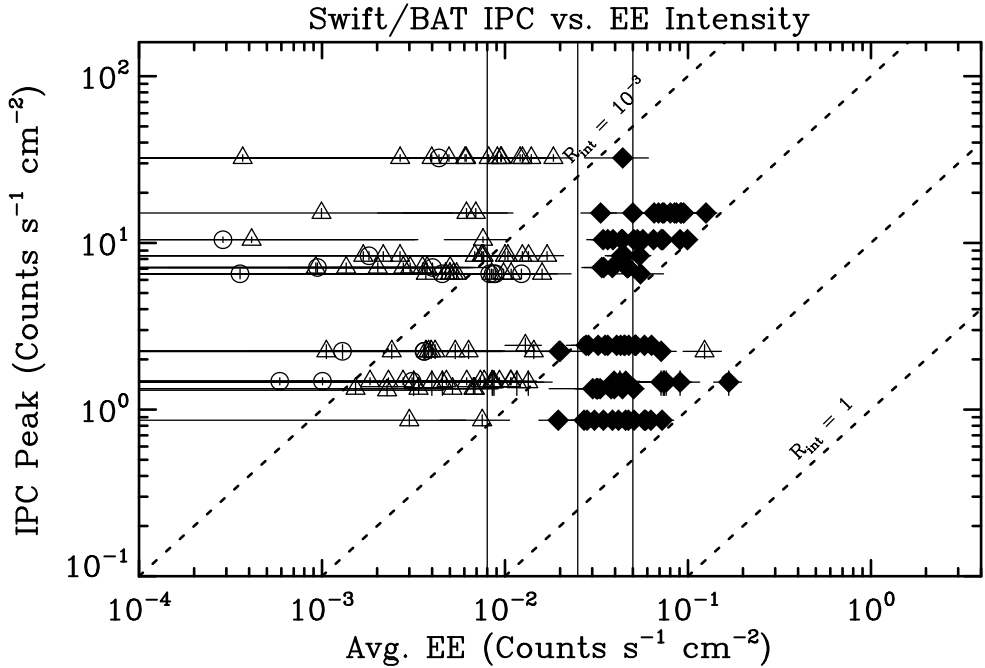

Fig. 2–Similar to Figure 1, but for 16 sets of simulations where the EE components were patterned after the BB representations of the twelve bursts with detected EE, and scaled to randomly uniformly populate the 0.008–0.05 counts cm$^{-2}$ s$^{-1}$ region in logarithm of EE average intensity. Some EE$_{avg}$ measurements fall above this region since higher intensity blocks are more readily detected, when in fact the simulated EE signal is multi-block. The simulated EE component is detected in 45% of the 192 cases – twice as frequently as for actual bursts.

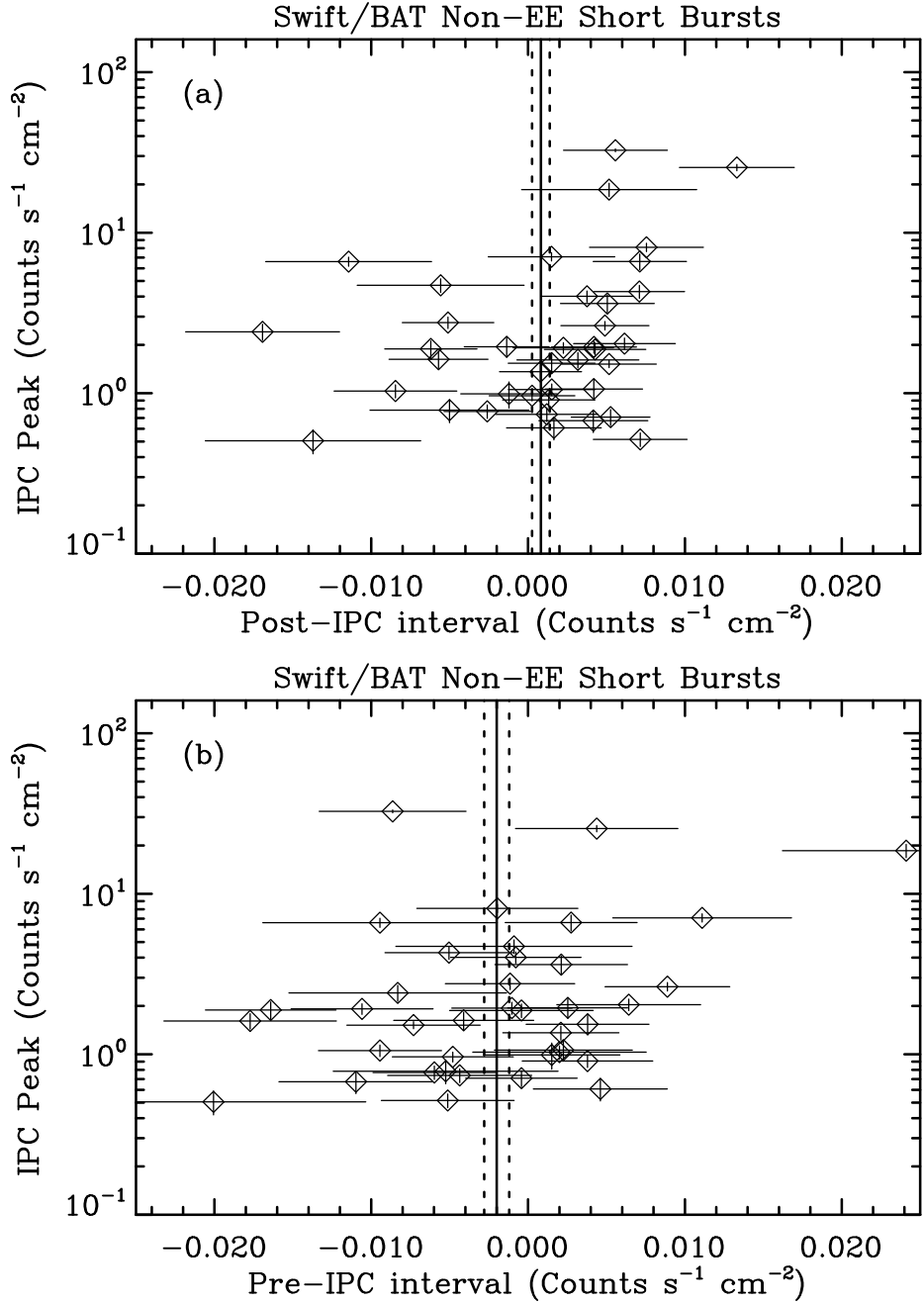

Fig. 3–Panel (a): The 1-$\sigma$ intensity upper limits in the post-IPC interval (~ 300–400 s) for the 39 bursts in Figure 1 without EE detections. Solid vertical line represents average of the upper limits, $8.2 \pm 5.5 \times 10^{-4}$ counts cm$^{-2}$ s$^{-1}$, a factor of 50 below the least intense EE detection; dashed lines are ± 1-$\sigma$ error bars. Highest positive and two most negative values are contaminated by very bright sources in BAT field of view, and would have smaller absolute values if the contaminations could be removed accurately (see text). Panel (b): As in Panel (a), but for the pre-IPC interval (~ 100–200 s), where no signal is expected prior to trigger time in short bursts.

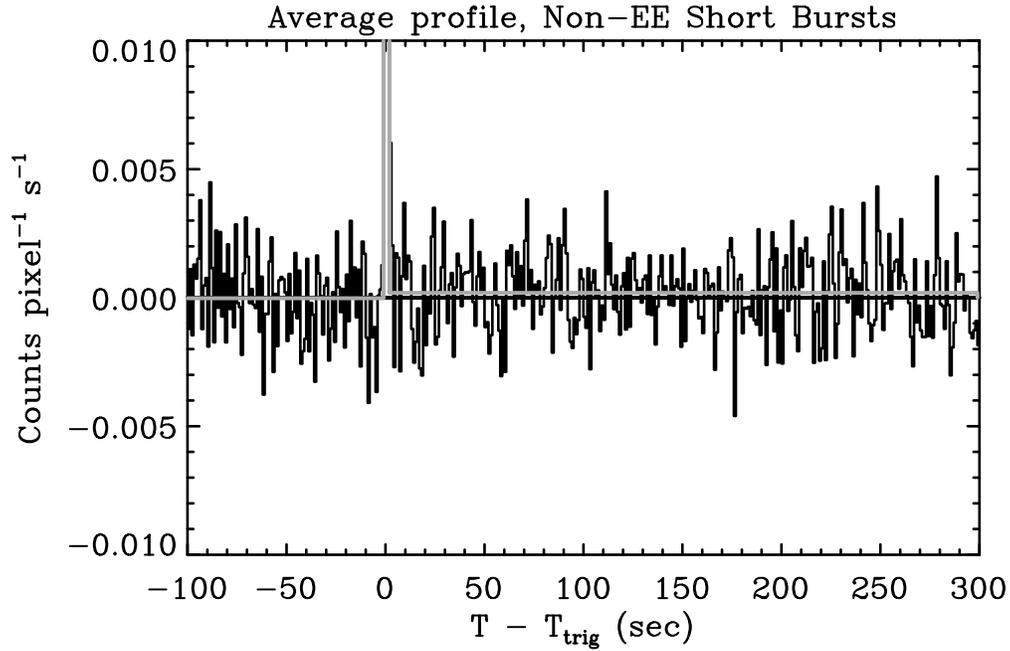

Fig. 4–The averaged time series for the 39 bursts with no detected EE component, truncated to the interval {–100, +300} s, with the individual time profiles aligned in registration at their trigger times. The BB representation of this average is overplotted in gray. Three blocks (–1 to +2 s) representing the average IPC are off scale (maximum = 0.14 counts pixel$^{-1}$ s$^{-1}$). The block representing the post IPC interval (2–300 s) is $1.9\pm1.0 \times 10^{-4}$ counts pixel$^{-1}$ s$^{-1}$ ($1.2\pm0.6 \times 10^{-3}$ counts cm$^{-2}$ s$^{-1}$).

## 4. DISCUSSION

For the twelve BAT bursts with a detected EE component the ratio of average EE intensity to IPC peak intensity, $R_{int}$, ranges over a factor of ~ 25, $R_{int}$ ~ $3 \times 10^{-3} - 8 \times 10^{-2}$. Whereas, for the 39 bursts with no EE detection, the block in Figure 4 representing the average post-IPC interval has an intensity of $1.2 \pm 0.6 \times 10^{-3}$ counts cm$^{-2}$ s$^{-1}$. Using the mid range of the IPC intensities in Figure 1, ~ 3 counts cm$^{-2}$ s$^{-1}$, the 2-$\sigma$ upper limit on $R_{int}$ for the non-EE portion of the sample is then $(1.2 + 2 \times 0.6) \times 10^{-3} / 3$ ~ $8 \times 10^{-4}$. These $R_{int}$ values for bursts with and without an EE component – and as shown in §3, the fact that an EE component could have been detected by the BAT in nearly half of the short burst sample, and the nearly unpopulated 99% – 1% detection zone in Figure 1 when source contamination is taken into account – all suggest that a physical threshold effect obtains near $R_{int}$ ~ few $\times$ $10^{-3}$, below which the EE component is not manifest in short bursts.

Of course, the lowest values for $R_{int}$ in Table 1 obtain for the bursts with brightest IPCs and no EE: GRBs 051221A, 060313, and 090510, with $R_{int}$ ranging ~ 2–5 $\times$ $10^{-4}$. Bursts with comparably bright IPC, but accompanied by EE (GRBs 050724, 061006, and 061210), have $R_{int}$ ~ 0.3–1 $\times$ $10^{-2}$. In fact, a glance at Figure 1 reveals that observed IPC intensity does not correlate with presence or absence of EE. This picture would probably not be too different upon translation to burst rest frames (that is, if redshifts were available for most of the sample), as the IPC and EE measures are in the same units.

If different progenitors gave rise to IPC-only and IPC+EE bursts, one might expect some differences in distributions of their IPC properties. As shown in Figure 5, the distributions of IPC $T_{1/e}$ durations (from Table 1) for the two BAT subsamples appear negligibly different, given the small number of bursts without EE (solid histogram). (A Kolmogorov-Smirnov test yields ±1-$\sigma$ confidence regions of [0.8, 2.7] on the relative stretch factor for the two distributions.) Also, for the BATSE short burst sample, the loci of fundamental burst characteristics – duration and spectral hardness (Kouveliotou et al. 1993) – were found to be indistinguishable, as shown in Figure 6 (from Norris & Gehrels 2009). In a separate work we will undertake an in-depth analysis of IPCs of short bursts, to determine if the two subsets exhibit significant differences.

Troja et al. (2008) suggested that the difference in host distances for IPC-only and IPC+EE short burst sources is commensurate with two progenitor types with different merger timescales, NS-NS and NS-BH systems. Our primary result from prompt emission evidence could in principle support the idea of distinct progenitors for the two short burst groups. From our work it appears that ~ three fourths of "short bursts" *are truly short*, of order ~ 1 s duration, and that one fourth manifests an additional mechanism that gives rise to the ~ 100-s EE component, which is part and parcel of the prompt emission. While the EE presence may be correlated with environmental properties, the latter probably do not directly cause the EE. Rather it could be that NS-BH mergers, occurring nearer the hosts, manifest the EE component due to a mechanism auxiliary to the IPC mechanism; whereas mergers of NS-NS binaries further from hosts, simply do not manifest the auxiliary mechanism. However, modeling indicates that there are far fewer NS-BH mergers than NS-NS mergers (e.g., Fryer & Kalogera 2001), perhaps insufficient numbers to account for the one fourth of short bursts with EE. The Troja et al. suggestion could still hold true if NS-BH systems are much more likely to make a GRB than a NS-NS merger.

Evidence emerged more than a quarter century ago for the dichotomous short/long classification of gamma-ray bursts (Mazets et al. 1981; Norris et al. 1984). Suggestions for any further fundamental divisions in this classification scheme have been less than robust. One of the primary considerations for a classification scheme is that it presage or reflect progenitor type, as does the fundamental division between short bursts (type I: e.g., NS-NS mergers) and long bursts (type II: stellar core collapse). Within the long-burst category, long-lag wide-pulse bursts, for instance, may qualify as a subset (Norris et al. 2005), but they probably still fall in the broad classification bin with core-collapse events – with a range in intrinsic physical properties (and a few extrinsic ones, e.g., observer viewing angle relative to jet) such as progenitor mass and/or angular momentum, commensurate with the observed diversity of long burst properties. The short burst dichotomy may reflect a similar situation – only one progenitor type – but with a threshold mechanism operating for the one quarter of short bursts with EE.

We would like to thank the anonymous referee for several important suggestions that improved the exposition of this work.

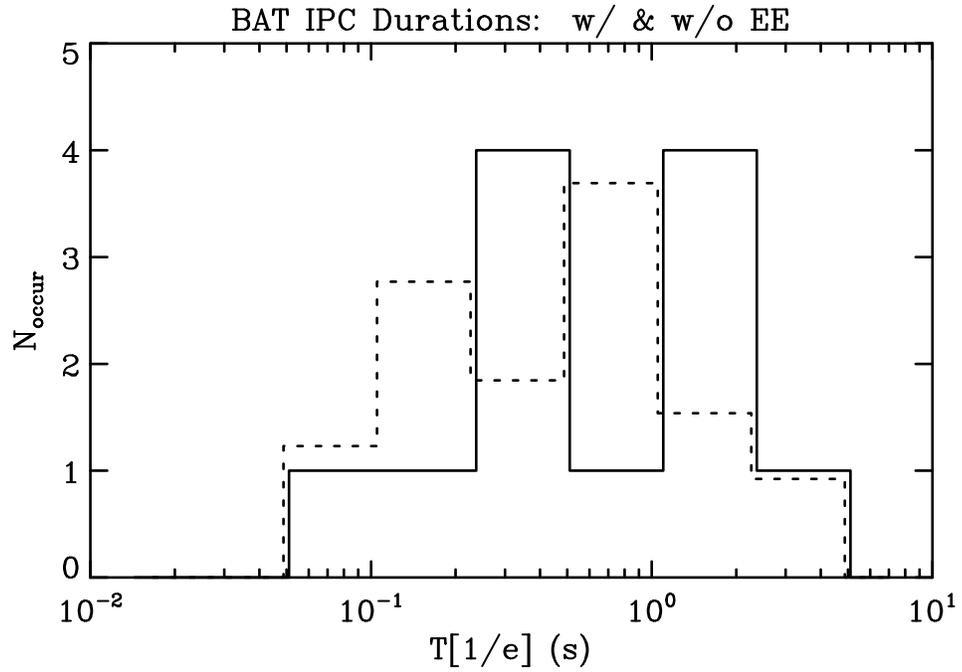

Fig. 5—$T_{1/e}$ duration histograms for the 51 bursts in Table 1, with (solid: 12 bursts) and without (dashed: 35 bursts) a detected EE component. Area of the latter histogram is normalized to that of the former.

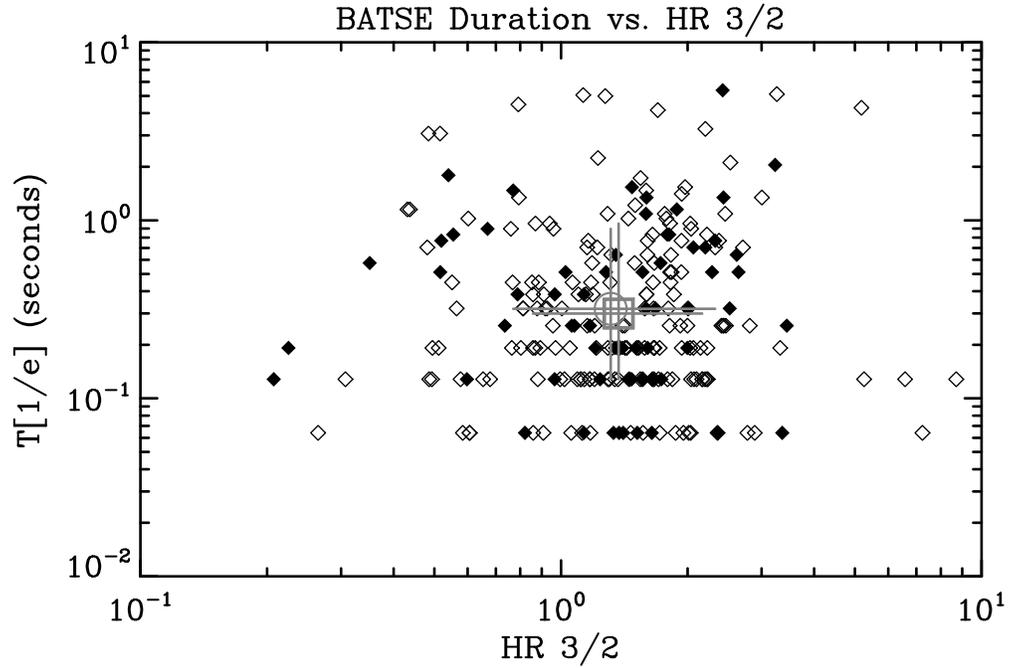

Fig. 6–For the IPC component, $T_{1/e}$ duration vs. hardness ratio (100–300 keV to 50–100 keV) for 256 BATSE short bursts, ~ one quarter of which showed evidence for an EE component. Filled (open) diamonds: bursts with (without) extended emission. The gray open circle (with EE) and box (without EE), with ±68% inclusion ranges, represent averages for the two subsets, statistically indistinguishable. Adapted from Norris & Gehrels (2009).

TABLE 1. Attributes of BAT Short Bursts with Extended Emission

| GRB Date | $T_{1/e}$ (s) | $IPC_{pk}$ (cm$^{-2}$s$^{-1}$) | $\varepsilon_{IPC}$ (cm$^{-2}$s$^{-1}$) | $T_{EE}$ (s) | $EE_{avg}$ (cm$^{-2}$s$^{-1}$) | $\varepsilon_{EE}$ (cm$^{-2}$s$^{-1}$) | $R_{int}$ | XRT[1] |
|---|---|---|---|---|---|---|---|---|
| 050202 | 0.128 | 2.413 | 0.277 | ----- | -0.0169 | 0.0049 | 0.0070 | N.Obs. |
| 050509B | 0.128 | 1.062 | 0.158 | ----- | 0.0042 | 0.0031 | 0.0040 | Y |
| 050724 | 0.384 | 8.376 | 0.496 | 104.4 | 0.0978 | 0.0080 | 0.0117 | Y |
| 050813 | 0.640 | 0.675 | 0.106 | ----- | 0.0042 | 0.0035 | 0.0062 | Y |
| 050906 | 0.128 | 0.991 | 0.184 | ----- | -0.0012 | 0.0031 | 0.0012 | Y |
| 050911 | 1.408 | 1.461 | 0.246 | 105.8 | 0.0529 | 0.0083 | 0.0362 | N.Obs. |
| 050925 | 0.128 | 6.605 | 0.435 | ----- | -0.0114 | 0.0053 | 0.0017 | N.D.P. |
| 051105A | 0.128 | 1.628 | 0.202 | ----- | -0.0057 | 0.0032 | 0.0035 | N.D.P. |
| 051210 | 1.408 | 0.515 | 0.050 | ----- | 0.0072 | 0.0030 | 0.0139 | Y |
| 051221A | 0.256 | 32.610 | 0.695 | ----- | 0.0056 | 0.0033 | 0.0002 | Y |
| 051227 | 0.832 | 0.863 | 0.067 | 119.1 | 0.0466 | 0.0051 | 0.0540 | Y |
| 060313 | 0.768 | 25.484 | 1.145 | ----- | 0.0133 | 0.0037 | 0.0005 | Y |
| 060502B | 0.128 | 2.755 | 0.227 | ----- | -0.0051 | 0.0029 | 0.0019 | Y |
| 060614 | 4.480 | 15.122 | 0.945 | 172.9 | 1.2354 | 0.0081 | 0.0817 | Y |
| 060801 | 0.576 | 1.055 | 0.087 | ----- | 0.0015 | 0.0028 | 0.0014 | Y |
| 061006 | 0.448 | 10.453 | 0.409 | 157.1 | 0.0667 | 0.0052 | 0.0064 | Y |
| 061201 | 0.640 | 6.616 | 0.744 | ----- | 0.0071 | 0.0030 | 0.0011 | Y |
| 061210 | 0.064 | 32.342 | 1.501 | 89.6 | 0.0911 | 0.0114 | 0.0028 | Y |
| 061217 | 0.256 | 1.362 | 0.136 | ----- | 0.0008 | 0.0026 | 0.0006 | Y |
| 070209 | 0.128 | 1.538 | 0.217 | ----- | 0.0015 | 0.0028 | 0.0010 | N.D.P. |
| 070429B | 0.640 | 1.888 | 0.230 | ----- | -0.0062 | 0.0029 | 0.0033 | Y |
| 070714A | 1.664 | 1.519 | 0.077 | ----- | 0.0052 | 0.0030 | 0.0034 | Y |
| 070714B | 1.152 | 7.116 | 0.464 | 71.1 | 0.0577 | 0.0074 | 0.0081 | Y |
| 070724A | 0.512 | 0.710 | 0.081 | ----- | 0.0053 | 0.0025 | 0.0074 | Y |
| 070729 | 1.088 | 0.770 | 0.074 | ----- | -0.0026 | 0.0028 | 0.0034 | Y |
| 070731 | 3.008 | 1.032 | 0.096 | ----- | -0.0085 | 0.0039 | 0.0082 | N.Obs. |
| 070809 | 1.472 | 0.963 | 0.068 | ----- | 0.0002 | 0.0027 | 0.0003 | Y |
| 070810B | 0.064 | 1.947 | 0.273 | ----- | -0.0014 | 0.0027 | 0.0007 | Y |
| 070923 | 0.064 | 4.701 | 0.460 | ----- | -0.0056 | 0.0053 | 0.0012 | N.Obs. |
| 071112B | 0.384 | 0.784 | 0.129 | ----- | -0.0050 | 0.0051 | 0.0064 | Y |
| 071227 | 1.088 | 1.334 | 0.100 | 106.6 | 0.0475 | 0.0068 | 0.0356 | Y |
| 080503 | 0.128 | 2.428 | 0.251 | 136.4 | 0.1562 | 0.0054 | 0.0643 | Y |
| 080702A | 0.576 | 0.609 | 0.097 | ----- | 0.0016 | 0.0030 | 0.0027 | Y |
| 080905A | 0.960 | 3.619 | 0.473 | ----- | 0.0051 | 0.0030 | 0.0014 | Y |
| 080919 | 0.704 | 0.908 | 0.080 | ----- | 0.0013 | 0.0029 | 0.0014 | Y |
| 081024A | 0.576 | 1.941 | 0.166 | ----- | 0.0042 | 0.0027 | 0.0022 | Y |
| 081226A | 0.512 | 1.874 | 0.180 | ----- | 0.0043 | 0.0032 | 0.0023 | Y |
| 090305A | 0.384 | 1.613 | 0.177 | ----- | 0.0032 | 0.0039 | 0.0020 | Y |
| 090426 | 1.408 | 1.917 | 0.145 | ----- | 0.0023 | 0.0032 | 0.0012 | Y |
| 090510 | 0.064 | 18.555 | 1.775 | ----- | 0.0052 | 0.0056 | 0.0003 | Y |
| 090515 | 0.064 | 4.299 | 0.394 | ----- | 0.0071 | 0.0029 | 0.0016 | Y |
| 090531B | 1.024 | 1.478 | 0.073 | 54.8 | 0.0912 | 0.0088 | 0.0617 | Y |
| 090607 | 2.560 | 0.739 | 0.073 | ----- | 0.0012 | 0.0032 | 0.0016 | Y |
| 090621B | 0.128 | 4.016 | 0.320 | ----- | 0.0038 | 0.0029 | 0.0009 | Y |
| 090715A | 0.320 | 6.524 | 0.433 | 63.7 | 0.1058 | 0.0194 | 0.0162 | N.Obs. |
| 090815C | 0.704 | 0.505 | 0.086 | ----- | -0.0137 | 0.0069 | 0.0272 | N.D.D. |
| 090916 | 0.448 | 2.236 | 0.215 | 67.4 | 0.0900 | 0.0157 | 0.0403 | Y |
| 091109B | 0.256 | 7.079 | 0.416 | ----- | 0.0015 | 0.0040 | 0.0002 | Y |
| 100117A | 0.384 | 2.038 | 0.169 | ----- | 0.0061 | 0.0032 | 0.0030 | Y |
| 100206A | 0.128 | 8.113 | 0.487 | ----- | 0.0075 | 0.0036 | 0.0009 | Y |
| 100213A | 2.432 | 2.638 | 0.160 | ----- | 0.0049 | 0.0028 | 0.0019 | Y |

---

[1] XRT legend: N.Obs. = Not Observed due to constraint. N.D.P. = Not Detected with Prompt slew. N.D.D. = Not Detected with Delayed observation.

APPENDIX

Fitness functions appropriate to a constant-rate model for a block of independently distributed time-tagged events (a recorded time for each photon), binned data, and time-to-spill data (fixed number of events per bin, variable time bins) for the Bayesian Block methodology were described earlier (Scargle 1998; the approximate algorithm in this paper is obsolete, having been replaced with the exact algorithm in Scargle et al. 2010). The results in the present paper were obtained with the much improved BB algorithm, a new fitness function for normally distributed data, and a procedure for choosing the algorithmic parameter (ncp$_{prior}$) characterizing the prior distribution of the number of blocks in the complete model. The details of these innovations are given in Scargle et al. (2010).

Here we describe this new fitness function, appropriate for the case of data consisting of a time sequence of measurements with Gaussian errors. But first we describe the adopted method of calibrating ncp$_{prior}$, as this is tailored to the current application. In particular, we adjust the parameter to yield a prescribed number (or fewer) of extraneous blocks in N$_{sim}$ realizations of 600 second time series like those of short bursts with extended emission.

The fundamental model assumes a constant signal over the time interval defining a block. The decision as to whether to allocate one, two, or more blocks (with different signal strengths) to a given time interval is based on evaluation of a fitness function for each of the cases. In addition, the locations of the divisions into blocks is based on maximizing the corresponding fitness function. The set of optimal divisions for a time series interval is global, i.e. based on all the information in the time series. However, data far from a particular block typically has much smaller influence than do nearby data. The optimal division algorithm, which takes into account the number of trial change points, was derived in Jackson et al. (2003). One of the advantages of the algorithm for estimating signals is that *it requires neither fixed time bins nor fixed counts per bin*, but can accommodate native data of any type.

Gaussianly distributed data are described by two data arrays, a signal ($s_i$) and associated error ($\sigma_i$) array, like the BAT mask-tagged data. The signal array is background subtracted. The details of the time bins for this data type – fixed or variable intervals, with or without data gaps – are irrelevant to evaluation of the fitness function; all the necessary information is contained in $\{s_i, \sigma_i\}$. All that matters is that the data arrays are sequentially ordered in time. The fitness function (an array of log-likelihoods) for each block in a set of blocks $\{B_k\}$ is

$$L_k = \Lambda_{k,1}^2/4\Lambda_{k,0} - \Lambda_{k,2} + \text{ncp}_{prior} \qquad (1)$$

where the auxiliary variables $\Lambda_{k,j}$ are defined as the following summations over all data points $\{s_i, \sigma_i\}$ lying within block k:

$$\Lambda_{k,0} = \tfrac{1}{2} \Sigma \, (1 / \sigma_i^2) \qquad (2)$$
$$\Lambda_{k,1} = -\Sigma \, s_i / \sigma_i^2 \qquad (3)$$
$$\Lambda_{k,2} = \tfrac{1}{2} \Sigma \, s_i^2 / \sigma_i^2 \, . \qquad (4)$$

Equation (1) is easily derived, under the single assumption of a normal distribution for the errors, by maximizing the block likelihood. The fitness function is manifestly invariant to a change of scale by factor $\alpha$ (e.g., a change of the units for the signal); that is, for $\{s_i, \sigma_i\} \rightarrow \{\alpha s_i, \alpha \sigma_i\}$ the value of the total fitness of $\{B_k\}$ does not change.

In equation (1), $ncp_{prior}$ is the single parameter in the adopted prior on the number of blocks. Its effect is in the sense that a larger value yields fewer block divisions. While it has an effect similar to that of a smoothing parameter, it does not smooth block edges; specifically it modulates the number of such edges in the representation of the whole time interval. This is the single free parameter in the BB formalism. As discussed in detail in Scargle et al. (2010), the optimal assignment for $ncp_{prior}$ depends on the time series. In fact, the assignment depends on interval length (in points) and collective signal-to-noise (S/N) ratio of the data points in that interval, and on total number of points in the time series. The obvious complication (which is addressable) is that in an arbitrary time series the signal can vary throughout the time profile – and hence the resulting blocks in the BB representation can vary in these two parameters {interval length, S/N ratio}.

Then the sufficiency requirement must be to adjust $ncp_{prior}$ so that those intervals in the time series with the least significant S/N ratio will have resulting blocks with high enough confidence to represent them, but no or very few extraneous blocks. That is, we desire BB representations with minimal occurrences of false positives (extraneous blocks) and false negatives (a block is not found where in fact it was simulated).

To illustrate the requirement for few false positives, it is instructive to consider a set of $N_{sim}$ simulated time series with single constant mean (signal level fixed, errors normally distributed) and fixed number of points. The expected BB representation is then one block per time series (there cannot be zero blocks – there will be no false negative in this case). For the present purposes all simulations detailed below are with $N_{pnts} = 600$ points (600 s), as are the majority of the BAT short burst time profiles analyzed in this work, when represented with 1-s binning. We explored the one-block case using sets of simulations with $N_{sim} = 10000$. For $N_{pnts} = 600$ with $ncp_{prior} \sim 6.5$ and unity S/N level, the resulting number of time series with one false positive is $\sim$ 200, or 2% – the canonical level we choose to tolerate. We varied $N_{pnts}$ to determine the dependence of $ncp_{prior}$ on time series length. For longer (shorter) time series $ncp_{prior}$ must be set lower (higher) to realize a false positive in only 2% of the 10000 simulations. The details of $ncp_{prior}$ dependences on S/N level and $N_{pnts}$ are thoroughly described in Scargle et al. (2010).

With the goal of assigning a global $ncp_{prior}$ value for this work, we examined a more realistic case similar to real-life short bursts with extended emission. We simulated a four-block model: pre-IPC background, IPC, EE, and post-EE background, where the respective interval lengths were 120, 1, $L_{EE}$, and 479 – $L_{EE}$ points (seconds). Four sets of simulations were performed, varying the length of the EE interval, $L_{EE}$ = 100, 50, 33, and 25 points, with commensurately decreasing post-EE background interval. We assigned unity S/N levels for the three long intervals, comparable to those in most BAT short burst time profiles in this study. For the IPC we assigned S/N level = 10, near the median for our burst sample. The IPC interval was always "detected" in the BB representation as one block: no false positive, no false negative in 10000 simulations. It is therefore evident that the IPC at the simulated S/N level is essentially like a "hard stop", in that the pre-IPC background will only rarely be represented by the BB algorithm to be longer than the simulated interval of 120 points (that is, in a rare simulation when the IPC is not represented by a block). But the pre-IPC background can be found to be shorter, since false positives (extraneous blocks) can occur in this interval. In contrast, the EE interval and the post-EE background interval share a change point, whose position will vary given the low S/N of these two components, and so the blocks found for these two components can be longer or shorter than their simulated values.

We accounted false positives and false negatives for multi-block models as follows. We assigned each extraneous block to the simulated block type within which the extraneous block's centroid occurred. Conversely, when no block's centroid fell within the interval of a simulated block, a false negative was accounted for that simulated block type.

Thus for our four-block model it is sufficient to adjust $ncp_{prior}$ only considering the three long intervals. With a value of $ncp_{prior}$ = 7.5, which we adopted for our analysis, the total number of false positives for the simulation set with $L_{EE}$ = 100 points is ~ 1%, with no false negative for any of the four block types. For $L_{EE}$ = 50 points the number of false positives is ~ 1.6%, and in only 0.3% simulations did the BB algorithm not find a block representing the extended emission interval. As $L_{EE}$ decreases to 33 and 25 points the number of false negatives increases more steeply, to 3.5% and then 12% – evidence of an approaching threshold below which the EE block will not be detected. Table 2 lists the numbers of false positives and false negatives for the four block types and for the four sets of simulations with different $L_{EE}$ values. The distributions of recovered block intervals are illustrated in Figure 7 for $L_{EE}$ = 100. For each block type the distribution is very narrowly peaked at the simulated value, with FWHMs of ~ 2–6 s (note that the ordinate is logarithmic). For the EE and post-EE background components the distributions are close to symmetric but with wide tails. For the pre-IPC component the distribution is almost a delta function with a low-level tail to shorter intervals. The IPC distribution is a delta function.

Clearly, prudent assignment of ncp$_{prior}$ for different types of time series requires understanding and prior characterization of the fundamental attributes of any signal(s) believed to be present.

In Figure 8 we illustrate three bursts with their BB representations overplotted. GRB 051221A was one of the brightest short bursts detected by *Swift*/BAT, but with no evidence for EE. GRBs 061006 and 090916 do have EE components, the former with, and the latter without a local minimum block between the IPC and EE.

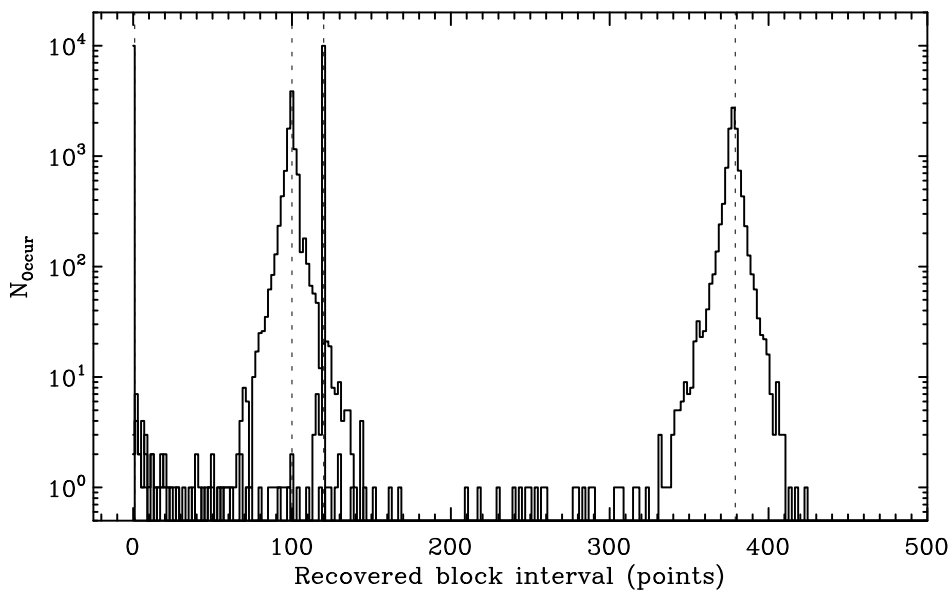

Fig. 7–Distributions of recovered block intervals for the four-block model for the case $L_{EE}$ = 100 points. Left to right peaks, the distributions are: IPC, EE, pre-IPC background, and post-EE background. IPC distribution is a delta function; pre-IPC distribution is nearly a delta function with low-level tail to shorter intervals; EE and post-EE distributions are nearly symmetric with wide tails. All distributions are very narrowly peaked at the simulated value.

TABLE 2. Number of False Positives & Negatives:  Four-Block Model

| Block Type | Simulated Length | FPs | FNs |
|---|---|---|---|
| Pre-IPC | 120 | 27 | 0 |
| IPC | 1 | 0 | 0 |
| EE | 100 | 28 | 0 |
| Post-EE | 379 | 43 | 0 |
| Pre-IPC | 120 | 42 | 0 |
| IPC | 1 | 0 | 0 |
| EE | 50 | 24 | 27 |
| Post-EE | 429 | 99 | 0 |
| Pre-IPC | 120 | 32 | 0 |
| IPC | 1 | 0 | 0 |
| EE | 33 | 12 | 347 |
| Post-EE | 446 | 137 | 0 |
| Pre-IPC | 120 | 41 | 0 |
| IPC | 1 | 0 | 0 |
| EE | 25 | 17 | 1220 |
| Post-EE | 454 | 216 | 0 |

For all cases:  $ncp_{prior} = 7.5$, and 10000 simulated time series.

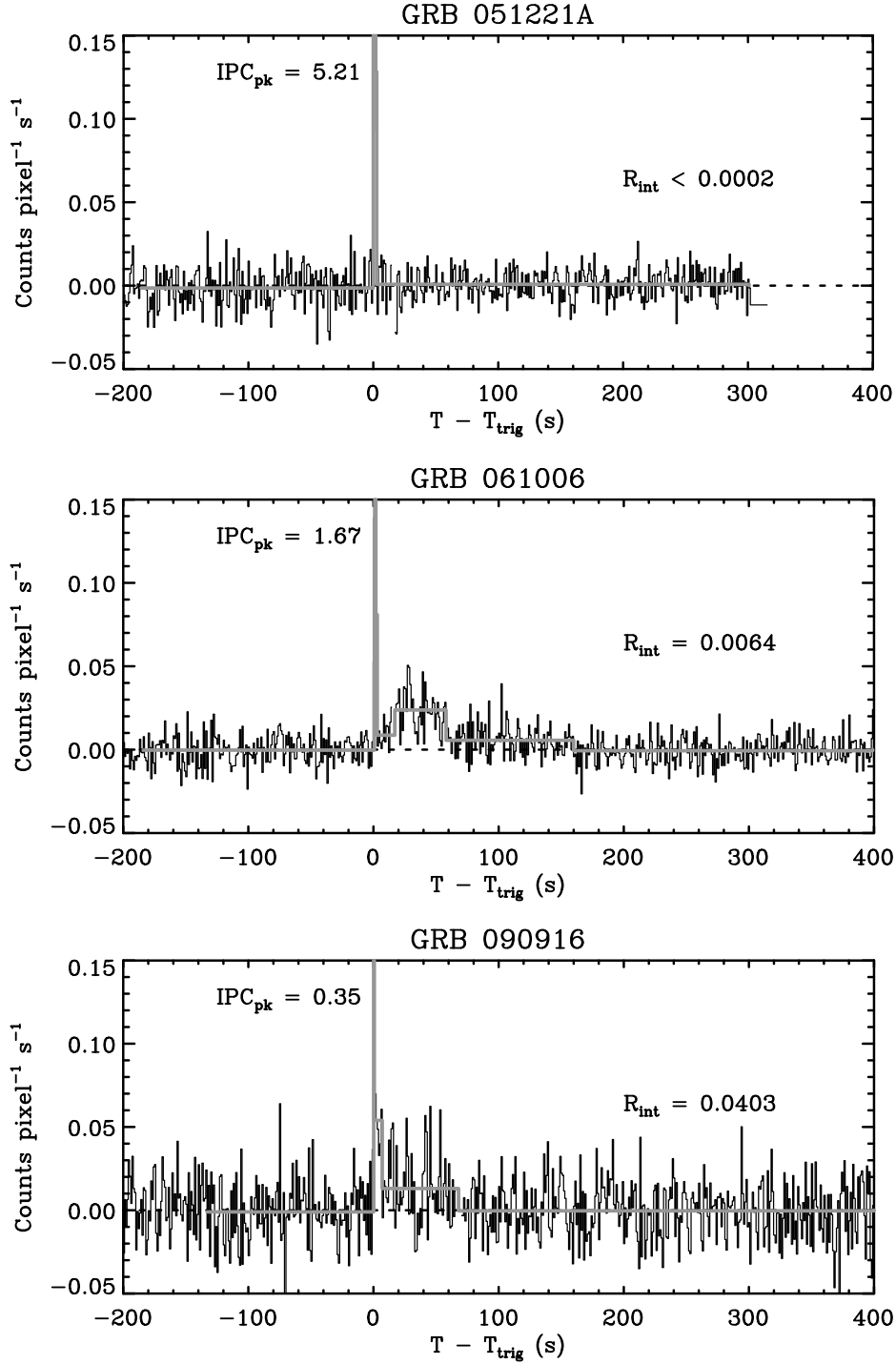

Fig. 8–Examples of BB representations overplotted on BAT 1-s mask-tagged data. Top panel: GRB 051221A, one of the brightest BAT short bursts, with no evidence for EE. Middle panel: GRB 061006, with a local minimum block between the IPC and EE. Bottom panel: GRB 090916 with no minimum between IPC and EE. Values for IPC peak intensities (off scale) in counts pixel$^{-1}$ s$^{-1}$ are annotated, along with $R_{int} = |$ IPC$_{pk}$ / EE$_{avg}$ $|$ values.